\documentclass{article}
\usepackage{latexsym}
\usepackage{amssymb}
\usepackage{pdfpages}
\usepackage{graphics}
\begin{document}
\newcommand{\ja}{Jakuba\ss a-Amundsen }
\newcommand{\bfx}{\mbox{\boldmath $x$}}
\newcommand{\bfq}{\mbox{\boldmath $q$}}
\newcommand{\bfnabla}{\mbox{\boldmath $\nabla$}}
\newcommand{\bfalpha}{\mbox{\boldmath $\alpha$}}
\newcommand{\bfA}{\mbox{\boldmath $A$}}
\newcommand{\bfe}{\mbox{\boldmath $e$}}
\newcommand{\bfn}{\mbox{\boldmath $n$}}
\newcommand{\bfW}{{\mbox{\boldmath $W$}_{\!\!rad}}}
\newcommand{\bfM}{\mbox{\boldmath $M$}}
\newcommand{\bfI}{\mbox{\boldmath $I$}}
\newcommand{\bfp}{\mbox{\boldmath $p$}}
\newcommand{\bfk}{\mbox{\boldmath $k$}}
\newcommand{\bfks}{\mbox{{\scriptsize \boldmath $k$}}}
\newcommand{\bfs}{\mbox{\boldmath $s$}_0}
\newcommand{\bfv}{\mbox{\boldmath $v$}}
\newcommand{\bfw}{\mbox{\boldmath $w$}}
\newcommand{\bfb}{\mbox{\boldmath $b$}}
\newcommand{\bfxi}{\mbox{\boldmath $\xi$}}
\newcommand{\bfzeta}{\mbox{\boldmath $\zeta$}}
\newcommand{\bfr}{\mbox{\boldmath $r$}}
\newcommand{\bfrs}{\mbox{{\scriptsize \boldmath $r$}}}

\title{\Large\bf The influence of vacuum polarization on the Sherman function during elastic electron-nucleus scattering}

%\author{D.~H.~Jakubassa-Amundsen} 
\author{D.~H.~Jakubassa-Amundsen \\
%Mathematics Institute, University of Munich, Theresienstrasse 39,\\ 80333 Munich, Germany}
Mathematics Institute, University of Munich,  Germany}

\date{}
%\date{\today}
\maketitle

\vspace{1cm}

\begin{abstract}  
The  change of the transverse spin asymmetry in the scattering of highly-relativistic spin-polarized electrons from nuclei by vacuum polarization effects
is calculated within the Uehling approximation. Results are provided for $1-100$ MeV electrons scattering from $^{31}$P, $^{64}$Zn, $^{152}$Sm, $^{197}$Au and $^{207,208}$Pb at backward angles.
\end{abstract}

Polarization phenomena during elastic electron-nucleus scattering \cite{Mo} provide deep insight into the nuclear structure because of their sensitivity to interference effects.
The measured spin asymmetries, being a relativistic efffect, are particularly large for heavy targets and backward scattering angles. Such conditions are accessible at the
electron accelerators in Darmstadt (S-DALINAC) and Mainz (MAMI).
In a pilot experiment a lead target was bombarded by 14 MeV electrons which were spin-polarized perpendicular to the beam axis.
By varying the scattering angle up to $172^\circ$ the influence of the finite nuclear size on the respective spin asymmetry, the so-called
Sherman function (also termed analyzing power), could be verified.
The accuracy of this measurement was about $3-5\%$.

Ambitious investigations require, however, high-precision measurements of the beam polarization and its control during the experiment.
An accuracy in the percent region or even below seems feasible in the near future due to a new generation of Mott and Compton polarimeters \cite{TAR,Ba}.
Such polarimeters will also be integrated into the chain of new optimized polarimeters designed for parity-violating electron scattering experiments \cite{Au2}.
For example, a precision measurement of the Weinberg angle $\theta_w$ is planned at the MESA facility in Mainz which soon will be under construction \cite{Au1}.

When the precision gets better than 1\%, quantum electrodynamic (QED) effects have to be considered in a calculation of the spin asymmetry at beam energies beyond the MeV region.
An accurate knowledge of the Sherman function does not only provide a prediction for experiment, but, once verified, it can be used to determine  the degree of beam polarization \cite{TAR}. Hence we
 address the question to which extent QED effects may change the Sherman function.
To lowest order in the fine-structure constant these QED corrections consist of the self-energy and the vacuum polarization (for their Feynman diagrams \cite{KK}, see Fig.1).
For heavy nuclei with a large charge number $Z$ the QED effects cannot be treated perturbatively, and  the field of the nucleus
has to be fully accounted for in any calculation of the electron propagator. Such calculations were performed for the 
QED corrections to electronic energy levels (see e.g. \cite{SM} for early work), radiative electron capture in ion-atom collisions \cite{Bei}
and radiative recombination in electron-atom collisions \cite{Sha}. For the latter, the changes in the cross section
were of the order of one percent (or below) at electron impact energies of $0.5-5$ MeV.

The present work is restricted to the consideration of  vacuum polarization. This provides a qualitative estimate of the QED corrections since the self-energy effects tend to be of opposite sign and of roughly twice the size of the vacuum-polarization effects \cite{Sha}.

\begin{minipage}[t]{9cm}
\includegraphics[angle=90,height=8cm]{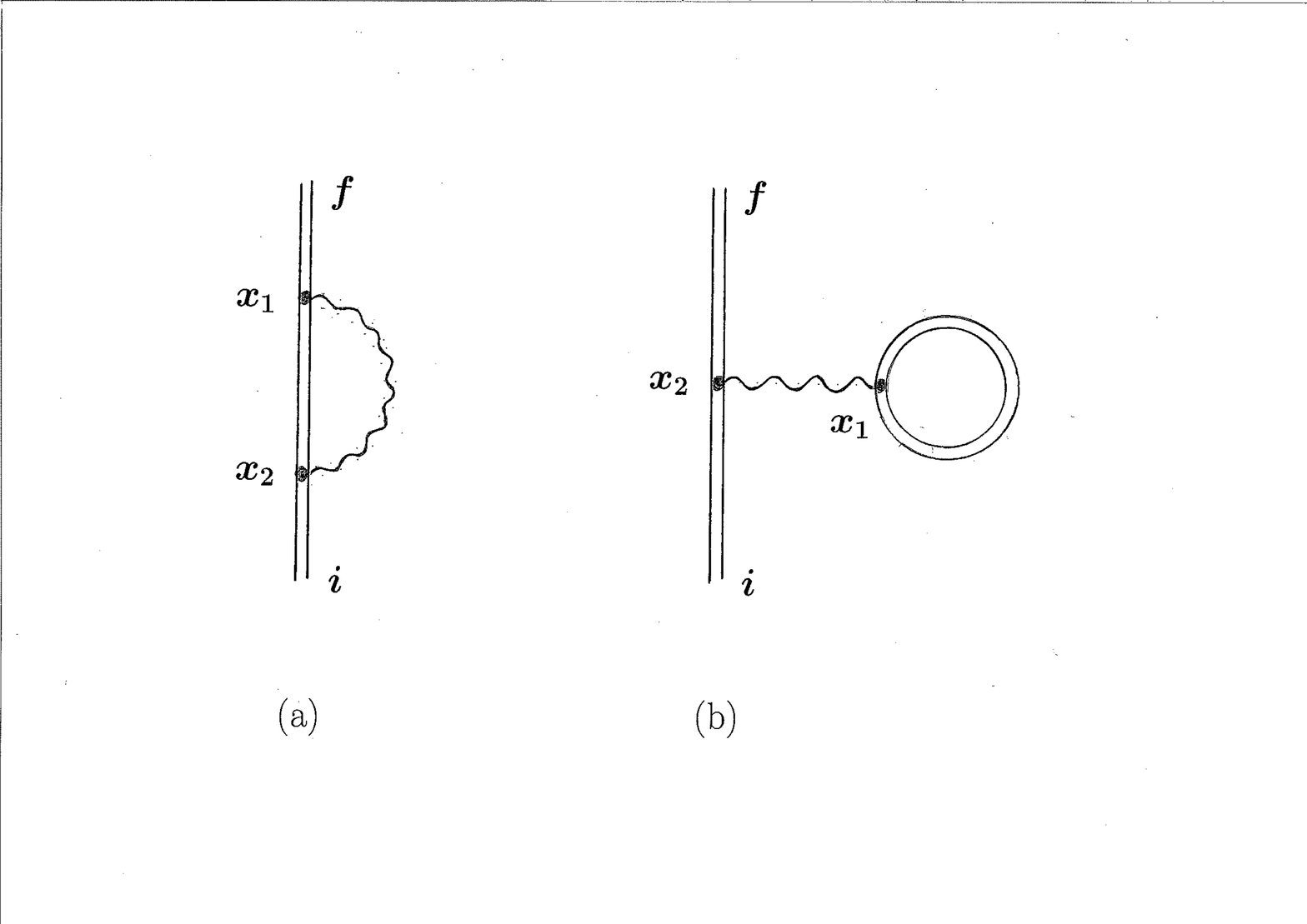}
\end{minipage}

Fig.1 QED corrections to elastic electron scattering.
(a) self-energy, (b) vacuum polarization. The double line indicates an electron in the field $V_T$ of the target nucleus, the wavy line a virtual photon.

\vspace{0.5cm}

The  transition matrix element for vacuum polarization during elastic scattering, as represented by the diagram in Fig.1b, is  (in atomic units, $\hbar=m=e=1$) given by \cite{BD,SM}
\begin{equation}\label{1}
S_{fi}\;=\;-\;\frac{1}{c^2}\int d^4x_1\;d^4x_2\sum_{\mu=0}^3 \left( \bar{\psi}_f(x_2)\;\gamma^\mu\;\psi_i(x_2)\right)\;D_F(x_2-x_1)\;\mbox{Tr}\,(\gamma_\mu S_F(x_1,x_1)),
\end{equation}
where  $\psi_i$ and $\psi_f$ are the dressed initial and final electronic states (i.e. scattering eigenstates to the nuclear potential $V_T$),
$D_F$ is the photon propagator, $\gamma_\mu$ are Dirac matrices 
and $S_F$ is the Feynman propagator of the dressed electronic states.
$S_{fi}$ can be rewritten in terms of an effective potential $U$,
\begin{equation}\label{2}
S_{fi}\;=\;-\;\frac{i}{c}\int d^4x_2\;\psi_f^+(x_2)\;U(x_2)\;\psi_i(x_2),
\end{equation}
where $U$ is a functional of $S_F$. The Uehling potential $U_e$ is obtained from this effective potential  by retaining only the term linear in $Z\alpha$ in an expansion of $S_F$
in terms of the nuclear potential strength and by renormalizing subsequently \cite{SM}. $U_e$ can be expressed by a simple integral \cite{Ueh,FR},
\begin{equation}\label{3}
U_e(\bfr) \;=\;-\;\frac{2\,Z}{3\pi c}\;\int d\bfr'\;\frac{\varrho(\bfr')}{|\bfr-\bfr'|}\;\chi_1(2c\,|\bfr-\bfr'|),
\end{equation}
where the function $\chi_n$ with $n=1$ is defined by
\begin{equation}\label{4}
\chi_n(x)\;=\;\int_1^\infty dt\;e^{-xt}\;\frac{1}{t^n}\;\left( 1\;+\;\frac{1}{2t^2}\right)\;\left(1\,-\,\frac{1}{t^2}\right)^{1/2},
\end{equation}
and  $Z\,\varrho(\bfr)$ is the nuclear charge density which generates $V_T$.
If $\varrho $ is spherically symmetric, (\ref{3}) reduces to \cite{KL}
\begin{equation}\label{5}
U_e(r)\;=\;-\;\frac{2}{3c^2}\;\frac{Z}{r}\int_0^\infty r'dr'\;\varrho(r')\;\left[ \chi_2(2c|r-r'|)\,-\,\chi_2(2c(r+r'))\right],
\end{equation}
which is small by a factor of $\alpha = 1/c$ as compared to the target field.
The function $\chi_2$ is given by (\ref{4}) with $n=2$. Alternatively, $\chi_1$ and $\chi_2$ can be parametrized
 in terms of rational functions of polynomials with coefficients tabulated in \cite{FR}. 

At asymptotically large distances compared to the nuclear radius $R_N$ (say, $r\gtrsim 50 \;R_N$) it is sufficient to keep the $\bfr'$-dependence on the rhs of (\ref{3})  only
 in $\varrho(\bfr')$, such that $U_e(r) $ reduces to the point-nucleus Uehling potential \cite{FR},
\begin{equation}\label{6}
U_e^\infty(r)\;=\;-\frac{2}{3c\pi}\;\frac{Z}{r}\;\chi_1(2cr),\qquad r\gg R_N.
\end{equation}

Rather than treating the modification of the elastic scattering by the Uehling potential perturbatively 
via adding (\ref{2}) (with $U$ replaced by $U_e$)
to the transition amplitude from potential scattering, the Dirac equation is solved
 in the combined potential $V_{VP}\,=V_T+U_e$,
\begin{equation}\label{7}
(-ic\bfalpha \bfnabla\,+\,\beta c^2\,+\,V_{VP}(\bfr))\;\psi_{VP}(\bfr)\;=\;E\;\psi_{VP}(\bfr),
\end{equation}
and the differential cross section $d\sigma/d\Omega$  for the elastic scattering of spin-polarized electrons from nuclei is  obtained from a phase shift analysis of the scattering states $\psi_{VP}(\bfr)$. 
In this way the Uehling part $U_e$ of the vacuum polarization potential is included to all orders \cite{Bei2,Bei}.
Note that in our approach the electron mass is retained but nuclear recoil effects are neglected.

For the solution of the Dirac equation the Fortran package RADIAL of Salvat et al \cite{Sa} was employed.
This was possible because the required initial condition on the potential ($\lim\limits_{r \to 0} r V_{VP}(r)< \infty$) is satisfied, since $\lim\limits_{r \to 0} \,rU_e(r)=0.$
Numerical details are described in \cite{Jaku12}.
In calculating $\psi_{VP}$ we have disregarded any screening by the bound target electrons since such effects do not play a role at the high momentum transfers considered here.
Magnetic scattering (from nuclei carrying spin \cite{Gro}) is also not  included in our estimate of the QED effects.
Its influence on the spin asymmetry depends strongly on the nuclear species \cite{Jaku14}  and is  small for high-$Z$ nuclei at low energy (see below).

The spin asymmetry $S$ for electrons spin-polarized perpendicular to the scattering plane (defined by the momenta of incoming and outgoing electron) 
 and scattering from an arbitrary potential $V$  is defined as the relative cross section difference when the spin of the beam electron is flipped (see e.g. \cite{Mo,Jaku12}),
\begin{equation}\label{8}
S(V)\;=\;\frac{d\sigma/d\Omega(\uparrow)\,-\,d\sigma/d\Omega(\downarrow)}{d\sigma/d\Omega(\uparrow)\,+\,d\sigma /d\Omega(\downarrow)}.
\end{equation}

Its change $d S$  by the influence of the Uehling potential is obtained from
\begin{equation}\label{9}
d S\;=\;\frac{S(V_{VP})}{S(V_T)}\;-1
\end{equation}
where $S(V_T)$ is calculated from the solution of (\ref{7}) with $U_e$ omitted.

We have investigated the backward scattering of electrons from the spin-zero nuclei $^{64}$Zn $\,(Z=30)$, $^{152}$Sm $\,(Z=62$) and $^{208}$Pb $\,(Z=82)$ as well as from the spin-$\frac12$ nuclei $^{31}$P ($Z=15$) and $^{207}$Pb. 
The respective potentials $V_T$  and $U_e$  were obtained  from the nuclear charge density which exists in terms of   a Fourier-Bessel expansion \cite{VJ}.
Also  $^{197}$Au $\,(Z=79)$, a favourite target for future experiments, was considered.
For this nucleus a two-parameter Fermi charge density was used \cite{VJ}.
The accuracy of our estimates for $d S$ is about 1\% below 40 MeV, but may deteriorate to 10\% or more near 100 MeV. The reason is that at high collision energies and backward angles the Dirac equation has to be solved numerically up to very large radial distances $(r>100\; R_N)$ and very high angular momenta ($l>100$) when $U_e$ is included,
in order to achieve sufficient accuracy in $S$.

For $^{197}$Au results for $S(V_T)$ and for $dS$ are shown in Fig.2  as a function of beam energy $E_{kin} =E-c^2$ at a scattering angle of $\theta=164^\circ$ which corresponds to the experimental set-up at the MAMI accelerator  for the planned  measurements of the beam  polarization  \cite{TAR}.
Since $U_e$ is negative  like the nuclear potential, the inclusion of $U_e$ leads to an enhancement of both the cross section and the spin asymmetry (in the sense that $S$ is more negative at the backmost angles if $U_e$ is included),
such that $d S >0$ 
 irrespective of $Z$ as long as the collision energy  is well below  50 MeV.
The enhancement of the cross section and of $d S$ increases monotonically with $E_{kin}$ for moderate energies as is generally expected for the QED effects \cite{Bei,Sha}.
For the Mott polarimeter operating at 3.5 MeV \cite{TAR}, $dS=0.36\%$. This results from $S(V_{VP})=-0.45408 $  and $S(V_T)= -0.45247$ (Fig.2b) according to (\ref{9}).

\begin{minipage}[t]{7cm}
\includegraphics[height=7cm]{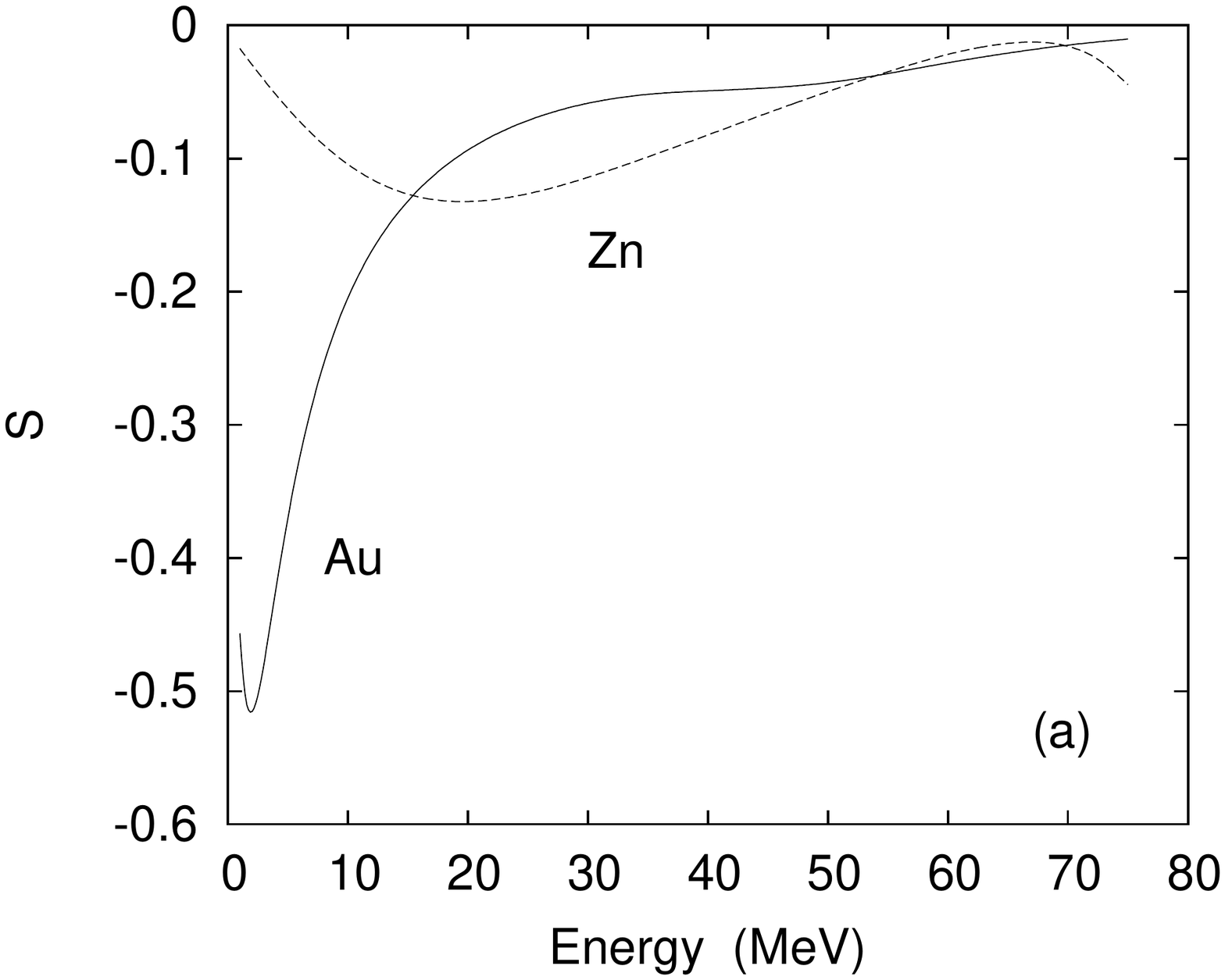}
\end{minipage}
\ \hspace{1cm} \
\begin{minipage}[t]{7cm}
\includegraphics[height=7cm]{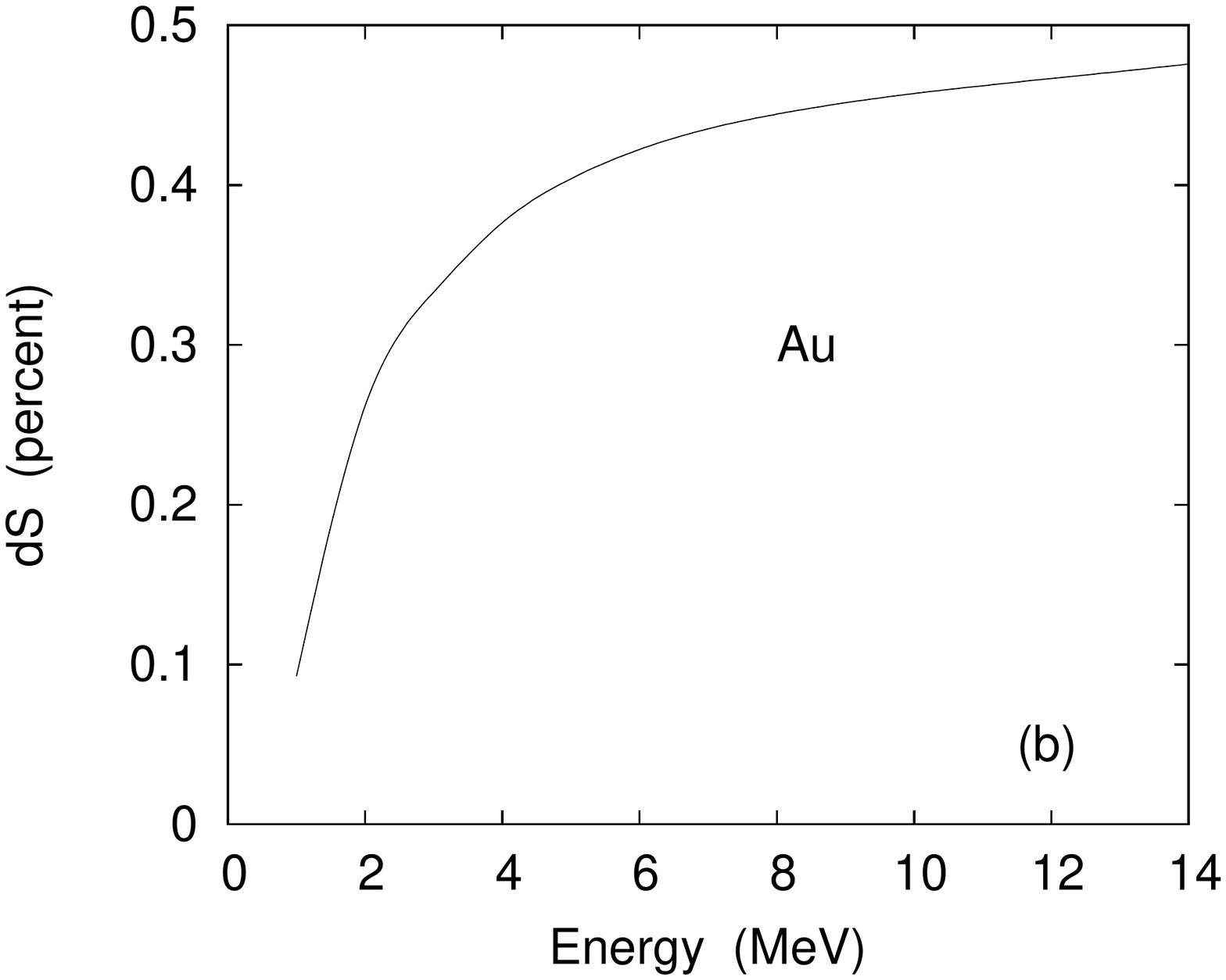}
\end{minipage}

Fig.2. (a) Spin asymmetry $S(V_T)$ from e + $^{197}$Au collisions at $\theta=164^\circ$ (full line) and from e + $^{64}$Zn at $178^\circ$ (dashed line) as a function of beam energy $E_{kin}$. (b) Change $dS$ for e + $^{197}$Au at $164^\circ$ according to (\ref{9}).

\vspace{0.5cm}

Fig.3 displays the change of spin asymmetry for the $^{152}$Sm and $^{207}$Pb targets, but now at a backmost scattering angle ($178^\circ$) which is accessible at the S-DALINAC set-up \cite{End}. $d S$ is shown in a large energy region where diffraction effects come into play.
Such effects are due to the finite nuclear size and arise when the electrons reach or penetrate the nuclear surface.
These structures occur not only in the cross section and in the spin asymmetry (see e.g. \cite{YRW,CH}, or Fig.2a for $^{64}$Zn) but, as seen from Fig.3, they also modulate the vacuum-polarization induced changes of $S$.
In particular, $d S$ shows oscillations with $E_{kin}$ in the same way as does $S(V_T)$ which is displayed in \cite{Jaku12} for $^{208}$Pb.
It should be noted, however, that $d S$ as defined by (\ref{9}) largely overestimates the influene of  vacuum polarization if the diffraction effects lead to a spin asymmetry which approaches zero or changes sign.
Such spurious effects from a near-zero denominator in (\ref{9}) cause e.g. the steep rise of $d S$ for $^{152}$Sm near 90 MeV, and for $^{64}$Zn near 60 MeV (see Figs.2a and 4).

Also shown in Fig.3 is the influence of  magnetic scattering on $S$ for $^{207}$Pb, which is of opposite sign and results from the current interaction
between the scattering electron and a nucleus carrying spin.
The respective changes in the spin asymmetry (without vacuum polarization effects) are calculated within the distorted-wave Born approximation (DWBA) as described in \cite{Jaku14}.
The scattering from the strong nuclear potential in $^{207}$Pb largely dominates the magnetic scattering even at an  angle as large as $178^\circ$.
Therefore, the magnetic effects  only come into play at energies near and above 100 MeV.
For the spinless isotope $^{208}$Pb the change $d S$ due to vacuum polarization is the same as shown in Fig.3 (up to 1\%),
while magnetic scattering is absent.

\begin{minipage}[t]{9cm}
\includegraphics[height=8cm]{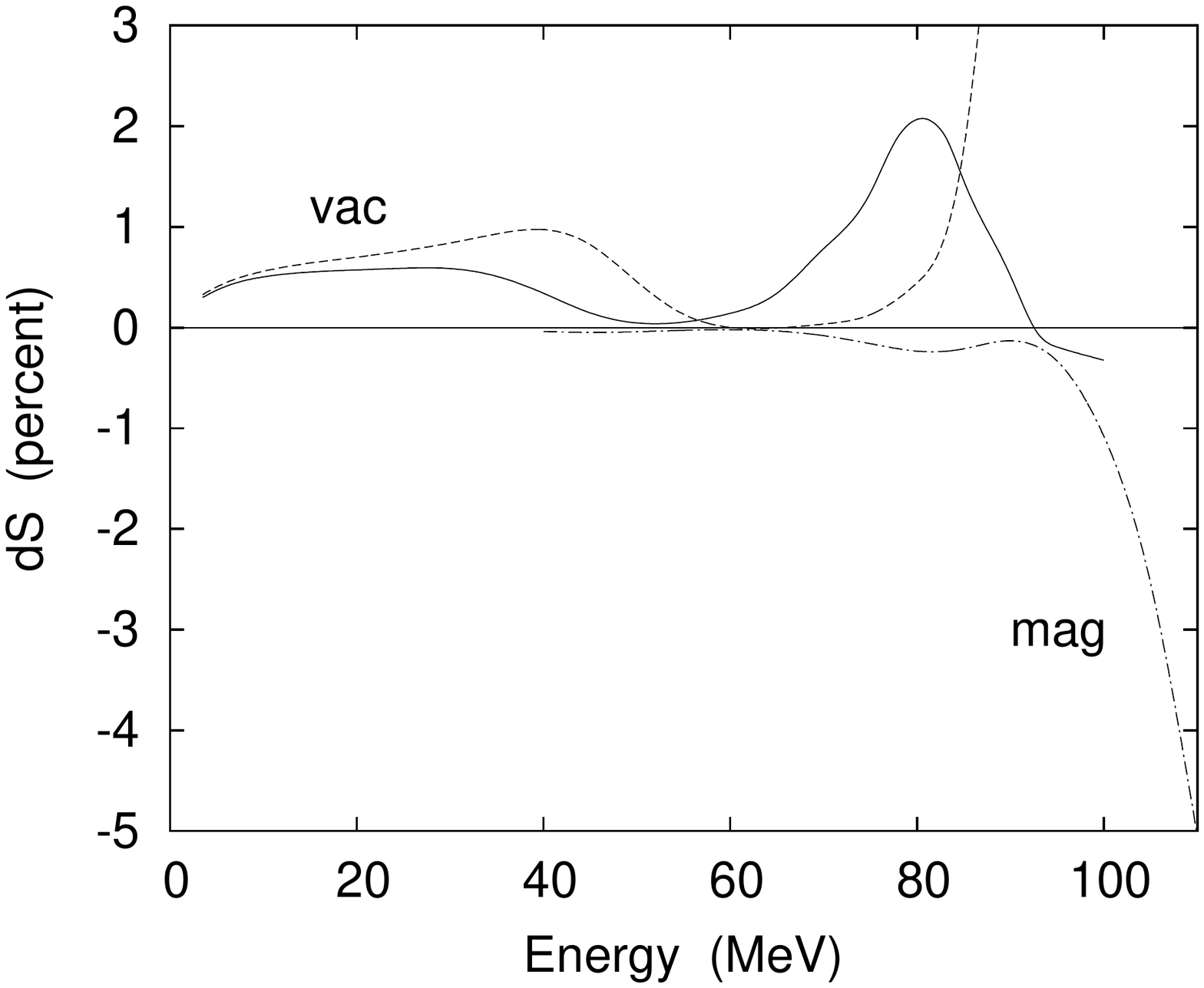}
\end{minipage}

Fig.3. Change $d S$ of the spin asymmetry from e + $^{152}$Sm and $^{207}$Pb collisions at $\theta=178^\circ$ as a function of collision energy $E_{kin}$. $^{207}$Pb: full line, effect from vacuum polarization as calculated from (\ref{9}); dash-dotted line, effect from magnetic scattering (with respect to $S(V_T)$), calculated within the DWBA \cite{Jaku14}.
$^{152}$Sm: dashed line, $d S$ from (\ref{9}).

\vspace{0.5cm}

Fig.4 shows the results for $^{31}$P and for $^{64}$Zn at $\theta=178^\circ$.
For these less extended nuclei the onset of the diffraction structures occurs at a higher energy, and the vacuum polarization effects increase smoothly 
 up to $E_{kin}\sim 50$ MeV.
However, for $^{31}$P, the  magnetic scattering plays a decisive role at much lower energies than in the case of $^{207}$Pb due to the larger magnetization current density and the weaker electric potential of this nucleus.
Thus, at energies beyond 40 MeV the 
 vacuum polarization effects are completely veiled.

\begin{minipage}[t]{9cm}
\includegraphics[height=8cm]{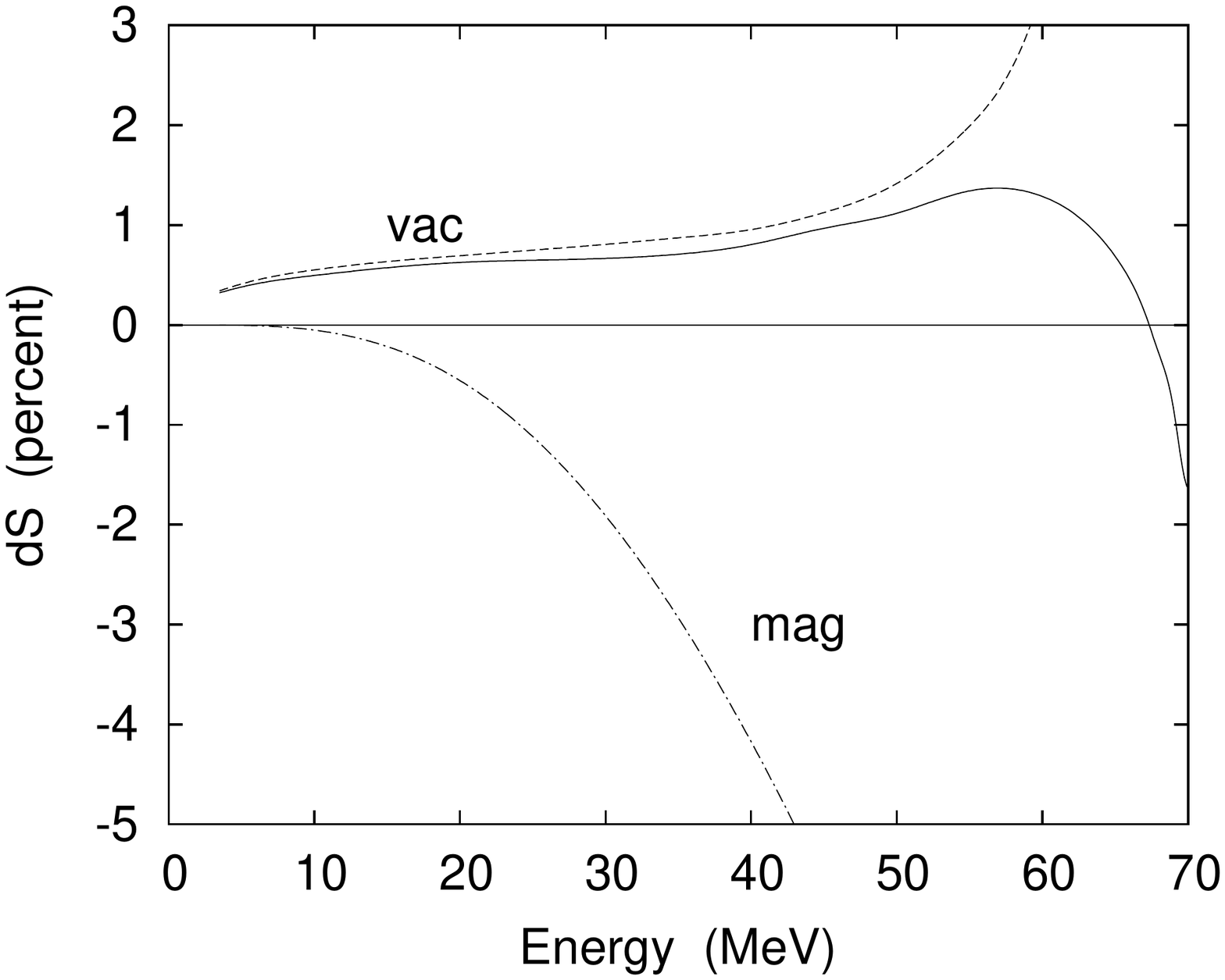}
\end{minipage}

Fig.4.  Change $d S$ of the spin asymmetry from e + $^{31}$P and $^{64}$Zn collisions at $\theta=178^\circ$ as a function of $E_{kin}$.
$d S$ as calculated from (\ref{9}): full line, $^{31}$P; dashed line, $^{64}$Zn.
Included is the change of $S$ by magnetic scattering for $^{31}$P \cite{Jaku13} (dash-dotted line).

\vspace{0.5cm}

The angular dependence of $d S$ is depicted in Fig.5 for the two collision energies 3.5 MeV and 30 MeV.
For the light target ($^{64}$Zn) as well as for the heavy target ($^{207}$Pb) the shift $d S \sim 0.3-0.4\%$ is approximately constant for $\theta \gtrsim 80^\circ$ at $E_{kin}=3.5 $ MeV.
In contrast, at the higher energy, $d S$ increases strongly with angle for both nuclei and flattens only beyond $130^\circ$.
The change in slope at the backmost angles is related to the steep minimum of $S(V_T)$ near $170^\circ$ (3.5 MeV), respectively at $178^\circ-179^\circ$ (30 MeV) \cite{Mo,Jaku13}.

\begin{minipage}[t]{9cm}
\includegraphics[height=8cm]{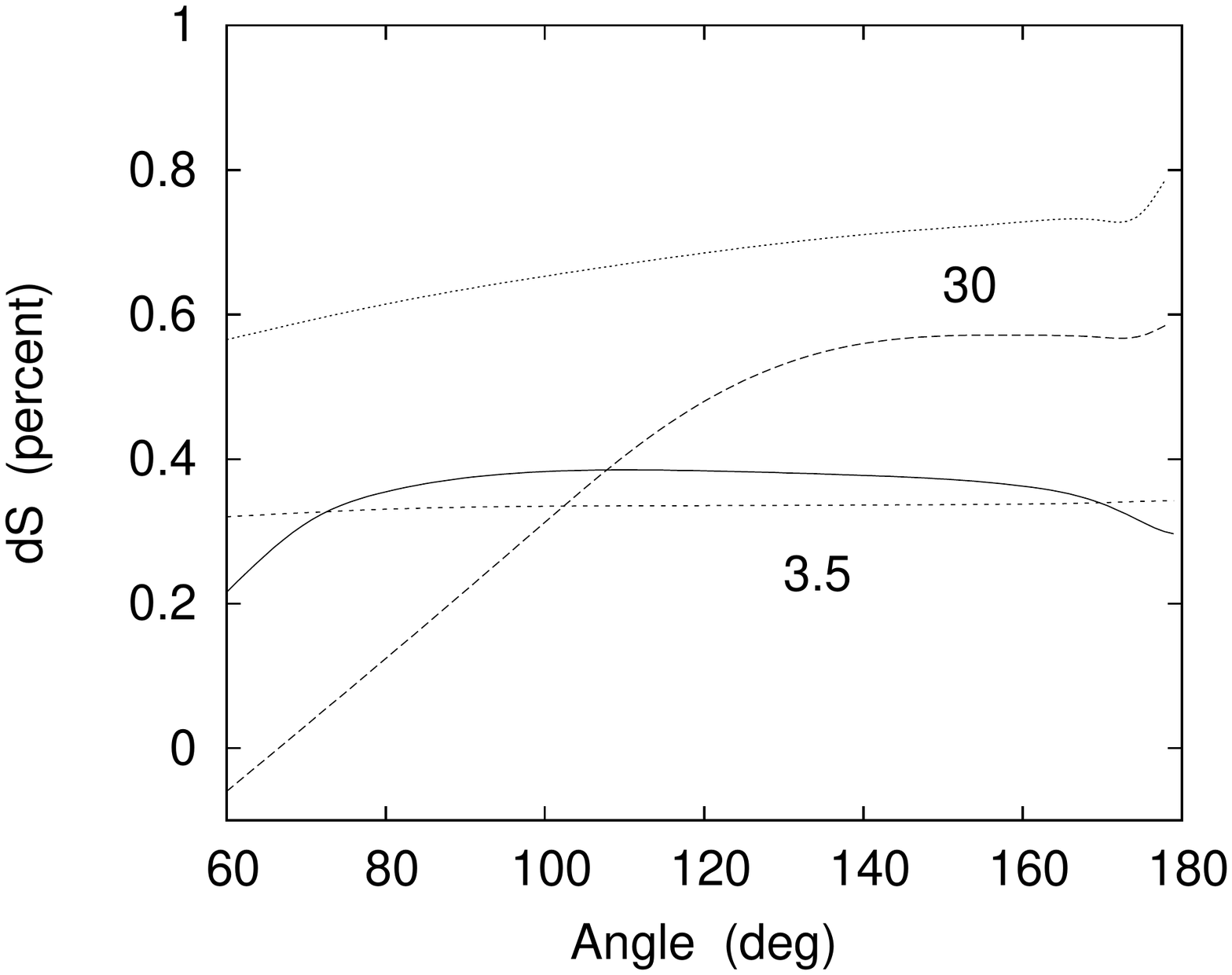}
\end{minipage}

Fig.5.  Change $d S$ of the spin asymmetry from e + $^{64}$Zn and $^{207}$Pb collisions at $E_{kin}=3.5$ and 30 MeV as a function of scattering angle $\theta$.
$^{207}$Pb: full line, 3.5 MeV; dashed line, 30 MeV.
$^{64}$Zn: short-dashed line, 3.5 MeV; dotted line, 30 MeV.

\vspace{0.5cm}

In conclusion, we have shown that at backward scattering angles the vacuum polarization contribution to the QED corrections affects the spin asymmetry  for a beam energy of 3.5 MeV  by $0.3 - 0.4\%$ for all nuclei investigated.
Up to $\sim 30$ MeV the change $d S$ of the spin asymmetry increases monotonically with $E_{kin}$ and is  weakly dependent on the nuclear species.
In the region $30-50$ MeV $d S$ is larger for the light nuclei ($^{31}$P, $^{64}$Zn),
while it decreases for the heavier nuclei at angles near $178^\circ$.
As a guideline, $d S \lesssim 1\%$ for energies near and below 40 MeV irrespective of the nuclear species. 

Beyond 40 MeV  modifications resulting from  the diffraction effects in electron scattering complicate the picture.
For  nuclei carrying spin the contribution of magnetic scattering also comes into play at the backmost angles, the more so, the lighter the nuclei.
Since the magnetization densities are not known to a sufficient accuracy it does not seem feasible to extract experimentally QED effects of the order of 1\% 
when magnetic scattering becomes important. Therefore, spin-0 nuclei are more appropriate candidates.
Alternatively, a decrease of the scattering angle $\theta$ leads to a moderate reduction of the vacuum polarization effects and to a disappearance of magnetic scattering.
However, since the spin asymmetry rapidly decreases when $\theta$ is lowered precision measurements will  become very difficult.

%\vspace{1cm}
\pagebreak

\noindent{\large\bf Acknowledgments}

I would like to thank K.Aulenbacher for stimulating this project and for helpful comments. I am also grateful to V.A.Yerokhin for valuable discussions.

\vspace{1cm}
%\pagebreak

%{\Large\bf Figure captions}

%\noindent{Fig.1}\\

\end{document}